\documentclass[aps,prd,twocolumn,nofootinbib,superscriptaddress]{revtex4-2}  

\usepackage{graphicx}  
\usepackage{dcolumn}   
\usepackage{bm}        
\usepackage{amsmath,amssymb,amsxtra,bm,epsfig}   
\usepackage{float}
\usepackage[dvipsnames]{xcolor}
\usepackage{xspace}
\usepackage{cancel,soul,ulem}

\usepackage{tikz-feynman}
\usepackage{stmaryrd}
\tikzfeynmanset{compat=1.1.0}
\usetikzlibrary{arrows}
\tikzset{	
	vertex/.style={circle,draw, minimum size=1.5em},	
	edge/.style={->,> = latex'}	
}
\usepackage[bookmarks, breaklinks, colorlinks,urlcolor=blue, citecolor=red, 
linkcolor=blue]{hyperref}

\hypersetup{
	colorlinks=true,
	linkcolor=red,
	filecolor=magenta,
	urlcolor=RoyalBlue,
	citecolor=ForestGreen}

\usepackage[]{caption}

\newcommand{\be}{\begin{eqnarray*}}
	\newcommand{\ee}{\end{eqnarray*}}

\newcommand{\bee}{\begin{eqnarray}}
	\newcommand{\eee}{\end{eqnarray}}
\newcommand{\beeq}{\begin{equation}}
	\newcommand{\eeq}{\end{equation}}

\usepackage{xspace}

\newcommand{\ba}{\begin{array}}
	\newcommand{\ea}{\end{array}}
\newcommand{\bd}{\begin{displaymath}}
	\newcommand{\ed}{\end{displaymath}}
\newcommand{\besub}{\begin{subequations}}
	\newcommand{\eesub}{\end{subequations}}
\newcommand{\bea}{\begin{eqnarray}}
	\newcommand{\eea}{\end{eqnarray}}

\def\q2 {q^2}

\usepackage{tikz}
\usetikzlibrary{arrows,shapes}
\usetikzlibrary{trees}
\usetikzlibrary{matrix,arrows} 				
\usetikzlibrary{positioning}				
\usetikzlibrary{calc,through}				
\usetikzlibrary{decorations.pathreplacing}  
\usepackage{pgffor}							

\usetikzlibrary{decorations.pathmorphing}	
\usetikzlibrary{decorations.markings}
\tikzset{
	vector/.style={decorate, decoration={snake}, draw},
	provector/.style={decorate, decoration={snake,amplitude=2.5pt}, draw},
	antivector/.style={decorate, decoration={snake,amplitude=-2.5pt}, draw},
	fermion/.style={draw=black, postaction={decorate},
		decoration={markings,mark=at position .55 with {\arrow[draw=black]{>}}}},
	fermionbar/.style={draw=black, postaction={decorate},
		decoration={markings,mark=at position .55 with {\arrow[draw=black]{<}}}},
	fermionnoarrow/.style={draw=black},
	gluon/.style={decorate, draw=black,
		decoration={coil,amplitude=4pt, segment length=5pt}},
	scalar/.style={dashed,draw=black, postaction={decorate},
		decoration={markings,mark=at position .55 with {\arrow[draw=black]{>}}}},
	scalarbar/.style={dashed,draw=black, postaction={decorate},
		decoration={markings,mark=at position .55 with {\arrow[draw=black]{<}}}},
	scalarnoarrow/.style={dashed,draw=black},
	electron/.style={draw=black, postaction={decorate},
		decoration={markings,mark=at position .55 with {\arrow[draw=black]{>}}}},
	bigvector/.style={decorate, decoration={snake,amplitude=4pt}, draw},
}

\tikzstyle{block} = [draw, rectangle, 
minimum height=3em, minimum width=6em]

\begin{document}
\title{Leptogenesis from a Phase Transition in a Dynamical Vacuum}

\author{Dipendu Bhandari}
\email{dbhandari@iitg.ac.in}
\affiliation{Department of Physics, Indian Institute of Technology Guwahati, Assam-781039, India}

\author{Arghyajit Datta}
\email{arghyad053@gmail.com}
\affiliation{Center for Precision Neutrino Research, Chonnam National University, Gwangju 61186, Republic of Korea }
\affiliation{Department of Physics, Kyungpook National University, Daegu 41566, Republic of Korea}

\author{Arunansu Sil}
\email{asil@iitg.ac.in}
\affiliation{Department of Physics, Indian Institute of Technology Guwahati, Assam-781039, India}

\begin{abstract} 


We show that a phase transition may take place in the early Universe at a temperature $T_*$ resulting a temperature dependent mass for right handed neutrinos (RHN) which finally relaxes to a constant value after electroweak symmetry breaking (EWSB). As a result, a requisite amount of lepton asymmetry can be produced near $T_*$ satisfying the observed baryon asymmetry of the Universe via sphaleron process 
even when zero temperature masses of the RHNs fall below the electroweak scale enhancing the detection possibility of RHNs. Interestingly, the framework is also capable of predicting a primordial lepton asymmetry (generated 
after EWSB) as hinted by helium abundance measurements, indicating a correlation with early phase of leptogenesis.

\end{abstract}
\maketitle

The precise measurement of cosmological parameters by PLANCK~\cite{Aghanim:2018eyx} and prediction of the big-bang nucleosynthesis (BBN)~\cite{Cyburt:2015mya} point toward a visible Universe with an excess of baryonic matter over antimatter, quantified as $n_B/s = 8.72 \pm 0.06 \times 10^{-11}$~\cite{Aghanim:2018eyx} where $n_B(s)$ is baryon number(entropy) density. In order to address the origin of such an excess, among many possibilities, production of baryon asymmetry via leptogenesis~\cite{Fukugita:1986hr,Luty:1992un,Plumacher:1996kc,Covi:1996wh} from the out-of-equilibrium decay of heavy Standard Model (SM) singlet RHNs $N_i$ into SM lepton ($\ell_L^T = (\nu_L, e_L)$) and Higgs ($H$) doublets, turns out to be the most motivated one as it can simultaneously account for the origin of neutrino mass via Type-I seesaw~\cite{Minkowski:1977sc,GellMann:1980vs,Mohapatra:1979ia,Yanagida:1979as,Schechter:1980gr,Schechter:1981cv,Datta:2021elq}. Eventually this lepton asymmetry gets converted to baryon asymmetry partially by sphaleron process~\cite{Kuzmin:1985mm,Arnold:1987mh, Bochkarev:1987wf, Khlebnikov:1988sr,Arnold:1987zg}.

The Lagrangian relevant for Type-I seesaw (in the charged lepton and RHN mass diagonal bases) is:
\begin{align}
	-\mathcal{L}_{\rm{I}}= \bar{\ell}_{L_\alpha} (Y_\nu)_{\alpha i} \tilde{H} N_i +\frac12 \overline{N_i^c} M_i N_i +h.c.,
	\label{type-I}
\end{align}
where $i=1,2$ (for minimal scenario) and $\alpha=e,~\mu,~\tau$ in general. Due to the seesawing nature of the light neutrino masses ($m_{\nu} = v^2 Y_\nu M^{-1} Y^T_\nu/2$ with $M=$ diag$\llbracket M_i\rrbracket$ and $v$ = 246 GeV), these RHNs 
are naturally (without any fine-tuning within Yukawa entries) thought to be very heavy ones. However, such a speculation is against the possibility of finding them in experiments. Also, from the thermal leptogenesis point of view, the mass 
of the lightest RHN ($N_1$) is bounded by $M_1 \gtrsim 10^9$ GeV (known as Davidson-Ibarra (DI) bound \cite{Davidson:2002qv}) for hierarchical RHNs\footnote{For leptogenesis happening during an extended reheating period, this bound gets modified as shown in \cite{Datta:2022jic,Datta:2023pav}.}.
Alongside, with quasi-degenerate RHNs, the associated CP asymmetry can be resonantly enhanced~\cite{Flanz:1994yx,Flanz:1996fb, Pilaftsis:1997jf,Pilaftsis:2003gt} which may bring the scale of RHN mass down to TeV scale\footnote{Some alternative scenarios exist where instead of RHN decay, leptogenesis proceeds via Higgs decay around $T_{\rm{sp}}$ only~\cite{Hambye:2016sby,Hambye:2017elz} and via CP-violating sterile neutrino oscillations \cite{Akhmedov:1998qx,Asaka:2005pn}.}. In an endeavor to bring this RHN mass scale further down so as to enhance its detection possibility, the first obstacle one would face is related to the decoupling of sphaleron at $T_{\rm{sp}}$ as below this temperature the lepton asymmetry, even produced, can't be converted to baryon asymmetry. 

The ambiguity in specifying the RHN mass scale (small or large) is further amplified by noticing that the corresponding bare mass term in the Lagrangian is not protected by any symmetry and, in principle, may assume any value up to the Planck scale.   A compelling alternative would be to introduce 
a global $U(1)_{B-L}$ symmetry ~\cite{Buchmuller:1992qc,Buchmuller:1996pa} and a SM scalar singlet $\phi$ charged under it, having an interaction
\begin{align}
	-\mathcal{L}_{\rm{int}} = \frac{1}{2}\alpha_i \phi \overline{N_i^c} N_i,
	\label{phi-nn}
\end{align}
which respects $U(1)_{B-L}$. Once the symmetry is spontaneously broken by the vacuum expectation value ($vev$) of 
$\phi$ ($v_{\phi}$), a Majorana mass term for the RHN, $M_i=\alpha_i v_\phi $, follows.  

We further venture into this possibility observing that such a $vev$ may evolve dynamically over a period of time in the expanding (radiation dominated) Universe with temperature $T$ before settling to a constant value. It turns out that a temperature dependent behaviour of $v_{\phi}(T)$ can approximately be parametrised as 
\begin{equation}
	v_\phi (T) \simeq
	\begin{cases}
		0& \text{;} \quad T>T_*,\\
		A T^2 & \text{;}  \quad  T_{\rm{c}}<T \leq T_*,\\
		A T^2+Bv^2(T) & \text{;}  \quad T\leq T_{\rm{c}},
	\end{cases}
	\label{eq:vphi}
\end{equation}
where $A$ and $B$ are constants having inverse mass dimension, and $T_* (T_c)$ represents the onset of the $U(1)_{B-L}$ (electroweak) phase transition, assumed to be smaller than the reheating temperature $T_{\rm{RH}}$ 
of the Universe following inflation. While the origin of such temperature dependence is detailed in the  Appendix~\ref{ap:1}, we point out here that the realization comprises of a potential for 
$\phi$ resulting from finite temperature correction over the zero-temperature (ZT) potential, 
thanks to an effective $\phi H^{\dagger} H$ term introduced.   
This Higgs-portal term contributes also during electroweak phase-transition (at and below $T_*$) in the form of shift in $vev$ of $\phi$ by $Bv^2(T)$ term with $v(T)= \sqrt{{(\mu^2 - c_h T^2)}/{\lambda}}$ where $c_h=0.4$; $\mu$ and $\lambda$ represent the quadratic and quartic coupling of the SM Higgs respectively. The framework requires the $\phi$ field to be in thermal equilibrium, at least till the electroweak (EW) scale, indicating its mass to be below $T_c$.
	The  construction differs fundamentally from few recent considerations \cite{Fardon:2003eh,Bi:2003yr,Brookfield:2005bz,ChoeJo:2023ffp,Borah:2023god,ChoeJo:2023cnx} as we motivate such temperature dependence from phase transition (PT) and leptogenesis perspective.

\begin{figure}[t]
\includegraphics[width=3.1 in]{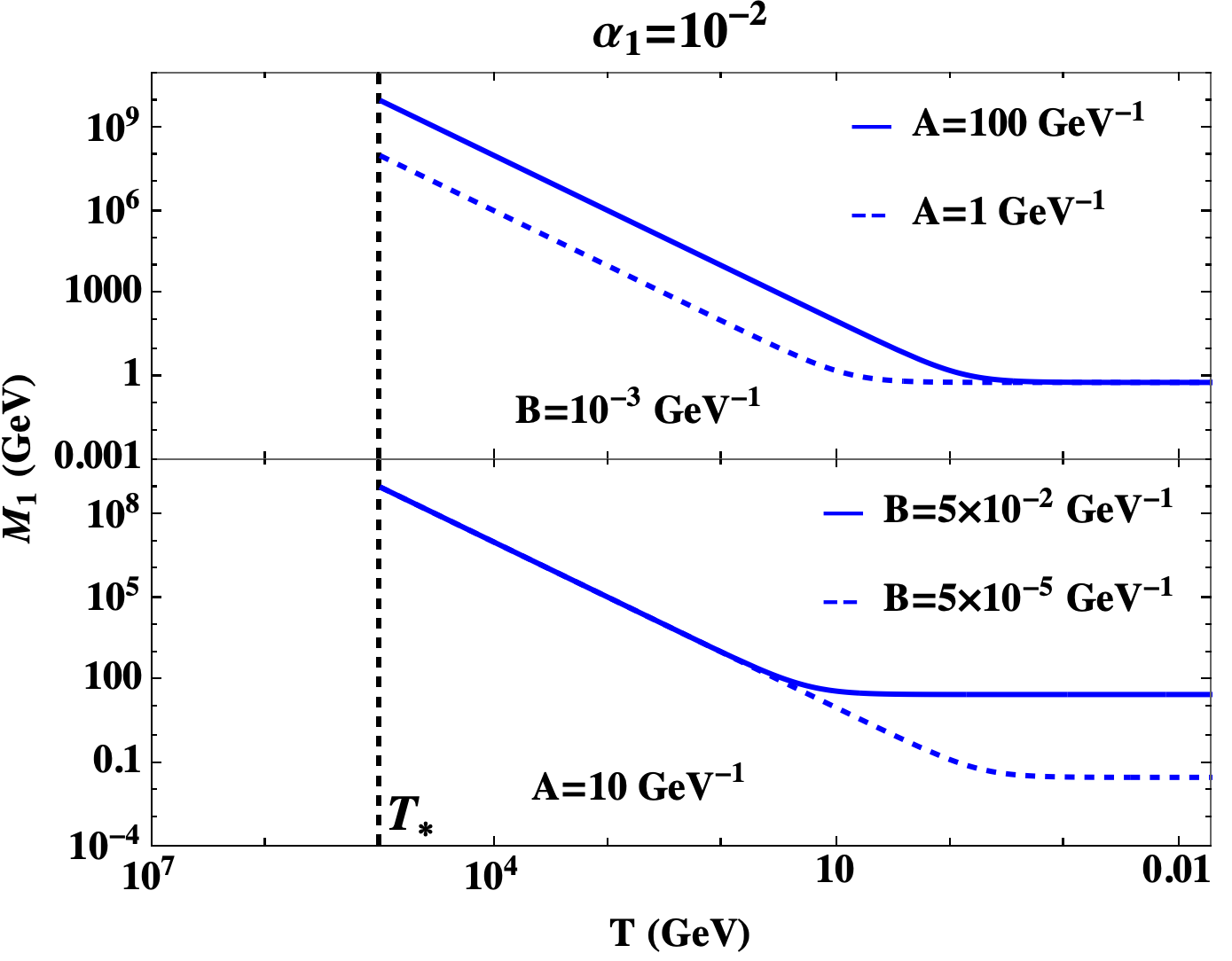}
\caption{RHN mass ($M_1$) variation with temperature.}
\label{fig:M1_T}
\end{figure}

In this work, for the first time, we demonstrate that such a temperature dependent $vev$ of $\phi$ not only carries a sizeable impact on the mass of the RHNs making it dynamical (T-dependent) above the EWSB, but also as a consequence of it, a viable leptogenesis scenario emerges even when the ZT mass of the RHNs ($M^0_i = \alpha_i B v^2$) remains smaller than the electroweak value. To be more specific, we plot the variation of mass of the lightest RHN as a function of temperature in Fig.~\ref{fig:M1_T}. We note that in the early Universe ($T > T_*$) with the $U(1)_{B-L}$ symmetry unbroken, RHNs are essentially massless. At $T_*$, mass of the RHN attains its maximum value and below this temperature, its mass decreases till a point (below $T_{\rm{c}}$) where it saturates to a value, much lower compared to $M_i(T_*)$. As seen from the upper panel of Fig.~\ref{fig:M1_T}, $M_1^0$ 
is 1 GeV (with which standard thermal leptogenesis is not possible) while its mass in the early Universe can be sufficiently high for  leptogenesis to happen.

In the early Universe above $T_* (> T_{\rm{c}})$, the RHNs being massless can be treated as relativistic particles and belong to the thermal bath having equilibrium number density $n^{\rm{eq}}_{N_i}$. However, as soon as $\langle \phi \rangle \neq 0$ at $T=T_*$, they suddenly become massive which depending upon the value of $A$ can even be greater than $T_*$. In this case with $M_i(T_*) > T_*$, both the RHNs decay immediately to SM lepton and Higgs doublets contributing 
to a finite amount of CP asymmetry, for hierarchical RHNs, as given by
\begin{align}
\varepsilon_{\ell} ^i=  \frac{1}{8 \pi (Y_{\nu}^{\dagger}Y_{\nu})_{ii}}\sum_{j\neq i}{\rm{Im}}\left[(Y_{\nu}^{\dagger} Y_{\nu})^2_{i j}\right] \mathbf{F}\left[\frac{M_j^2(T)}{M_i^2(T)}\right],
\label{eq:cp}
\end{align}
where $\mathbf{F}[x]= \sqrt{x}\left[1+\frac{1}{1-x}+(1+x)\ln\left(\frac{x}{1+x}\right)\right]$ is the loop function inclusive 
of both vertex and self-energy diagrams.   
The structure of $Y_{\nu}$ matrix can be extracted using Casas-Ibarra (CI) parametrization \cite{Casas:2001sr},
\begin{align}
Y_{\nu}=-i \frac{\sqrt{2}}{v} U D_{\sqrt{m}} \mathbf{R} D_{\sqrt{M^0}}\,,
\label{CI}
\end{align}
where $U$ is the Pontecorvo-Maki-Nakagawa-Sakata matrix~\cite{Zyla:2020zbs} which connects the flavor basis to the mass basis of light neutrinos. Here $ D_{\sqrt{M^0}}=\rm{diag}(\sqrt{M^0_1},\sqrt{M^0_2})$ and $ D_{\sqrt{m}}=\rm{diag}(\sqrt{m_1},\sqrt{m_2},\sqrt{m_3})$ denote the diagonal matrices containing the square root of ZT RHN masses and light neutrino masses respectively with $\mathbf{R}(\theta)$ representing a complex orthogonal matrix.


This CP asymmetry therefore results a $B-L$ asymmetry at $T_*$ that can approximately be given by 
\begin{align}
Y_{B-L}\simeq -\sum_i \varepsilon^i_{\ell} \frac{n_{N_i}(T_*)}{s}= -\frac{3}{4}\sum_i \varepsilon^i_{\ell}\frac{T_*}{M_i(T_*)}\,. 
\label{y_b-l}
\end{align}
Note that at $T_*$, the equilibrium number density of the RHNs are employed for $n_{N_i}(T_*)$ since immediately above $T_*$ the RHNs were massless and in thermal equilibrium, as already stated. We assume the instantaneous conversion of energy densities of both the RHNs into radiation, $\rho_{N_{i}} (T_*) = M_i (T_*) n^{\rm{eq}}_{N_i} \equiv \rho_R = (\pi^2/30) g_* T^4_*$. Note that $\varepsilon^i_{\ell}$ is effectively $T$ independent as 
the temperature in the argument $x$ of $\mathbf{F}[x]$ is cancelled. Therefore, the CP asymmetry turns out to be function of parameter $B$ while the factor $T_*/M_i(T_*)$ crucially depends 
upon $T_*$ and $A$ (apart from the $\alpha_i$ dependence). 


We need to finally evaluate how much of the above asymmetry would survive till $T_{\rm{sp}}$ beyond which conversion into baryon asymmetry won't be possible. It is interesting to note that during $T_{\rm sp}< T \leq T_*$, $M_i$ falls faster ($M_i\propto T^2$) compared to temperature. As a result, a situation may arise, depending on the choice of $A$, where $M_i$ may drop below $T$ before $T_{\rm sp}$ (or $T_{\rm{c}}$). Consequently, production of RHNs via inverse decay becomes possible, causing a complete erasure of the $B-L$ asymmetry produced earlier. So, from the survival point of view of the early $B-L$ asymmetry, one needs to maintain $M_i(T) > T$ at least till $T_{\rm{sp}}$ by choosing appropriate value of $A$ so as to convert it to baryon asymmetry by sphaleron process 
using $Y_B = (28/79) Y_{B-L}$ ~\cite{Harvey:1990qw} which holds for $T_{\rm{sp}} > T_c$.

\begin{figure}[!htb]
\includegraphics[width=3.1 in]{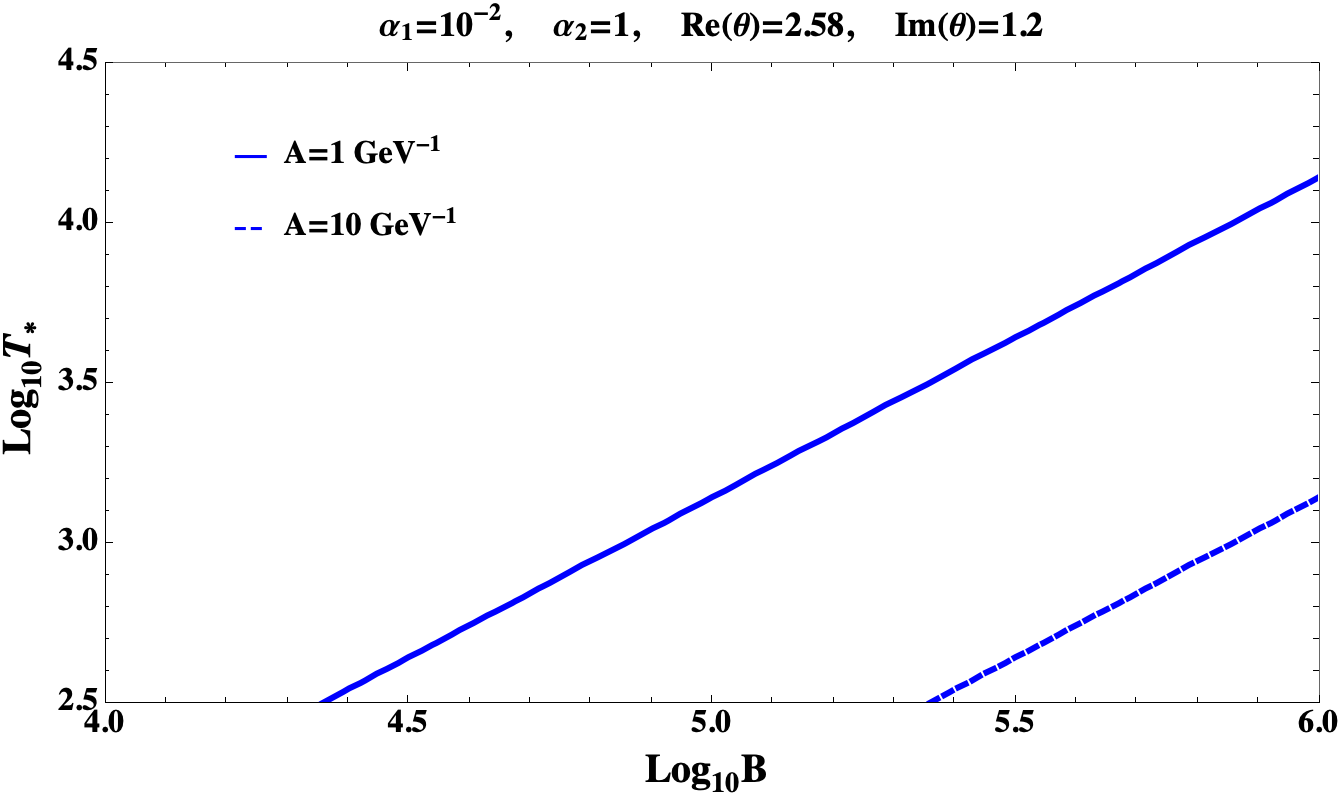}
\caption{Contour of $Y_B=8.72 \times 10^{-11}$ in $T_*$ - $B$ plane.}
\label{fig:contour}
\end{figure}
As evident from the discussion above, we have parameters $T_*,~A,~B$ and $\alpha_i$ involved in the study. A minimum value of $A$ is obtainable from $M_1(T) \geq T$ satisfaction at $T = T_{\rm{sp}}$ leading to $A \geq (\alpha_1 T_{\rm{sp}})^{-1}$. Furthermore, $T_*$ should maintain $T_* >T_{\rm sp}$ in order to have a non-zero baryon asymmetry, which leads to a minimum choice of $A$ as $\mathcal{O}(1)$ GeV$^{-1}$. 
Using the naive analytic estimate of Eq.~\eqref{y_b-l}, we provide a contour plot in Fig.~\ref{fig:contour} for $Y_B = 8.72 \times 10^{-11}$ in $T_* - B$ plane for some fixed choices of $A$ with $\alpha_1=10^{-2}$, $\alpha_2=1$ (for hierarchical RHNs). Note that, $Y_{\nu}$ is evaluated from Eq.~\eqref{CI} using the best-fit values of neutrino oscillation data \cite{Esteban:2020cvm} corresponding to each $B$ value. As observed, the minimum value of $B$ so as to produce the correct amount of baryon asymmetry (satisfying light neutrino mass) is found to be $\mathcal{O}(10^4)$ ${\rm GeV^{-1}}$ for $A = 1$ GeV$^{-1}$, which leads to the bare ZT mass of the lightest RHN as $M_{1}^0 \simeq \mathcal{O}(10^7)$ GeV.  This acts as the lowest possible RHN mass scale compatible with the scenario we have. With a further increase in $A$, a higher value of $B$ and obviously a heavier $M_{1}^0$ would follow to have correct $Y_B$. Note that such minimum $M_1^0$ apparently evades the DI bound for thermal leptogenesis. This observation is actually connected to state of RHNs across $T_*$ that is somewhat intermediate between completely thermal and non-thermal scenarios. It follows from the fact that the RHNs although was part of the thermal bath  for $T > T_{*}$, due to sudden initiation of the PT of $\phi$ at $T_*$, found themselves massive as well as non-relativistic (since $M_1(T_*) > T_*$) instantly. As a consequence, they immediately decay producing the lepton asymmetry mimicking non-thermal leptogenesis. However, due to the absence of a parent field/source for such non-thermal production, one can't assign it as purely non-thermal leptogenesis too.

The above finding that the early leptogenesis may even occur with ZT RHN mass as low as $10^7$ GeV motivates us to go one step further and look for a scenario where RHNs of even smaller ZT mass be able to produce the required amount of $B-L$ asymmetry. A natural possibility in this direction would be to investigate the resonant leptogenesis. In that case, a resonant enhancement of CP asymmetry $\sim \mathcal{O}(1)$ may occur at $T_*$ with $N_{1,2}$ being quasi degenerate having mass $\sim M(T)$ satisfying the resonance condition at $T_*$ as $\Delta M (T) = M_2 (T) - M_1(T) \sim \Gamma_{N_1} (T)/2$ leading to a large amount of $B-L$ asymmetry generation.
The dominant contribution to the CP asymmetry comes from the interference between tree level decay and one-loop self-energy correction \cite{Pilaftsis:1997jf,Pilaftsis:2003gt}, expressed as
(above $T_{\rm{sp}}$) 
\begin{align}
\varepsilon_{\ell}^i=\sum_{j\neq i}\frac{\text{Im}(Y_\nu^\dagger Y_\nu)_{ij}^2}{(Y_\nu^\dagger Y_\nu)_{ii}(Y_\nu^\dagger Y_\nu)_{jj}} \frac{\left[M_i^2 -M_j^2 \right] M_i \Gamma_{N_j}}{\left[M_i^2 - M_j^2\right]^2+M_i^2\Gamma_{N_j}^2},
\end{align} 
where $i,j=1,2$ and $T$ dependence of masses $M_i$ are kept implicit. 

To observe the impact of resonant leptogenesis with   $M_i^0$ below EW scale, a smaller value of $B$ becomes inevitable, obvious from the bottom panel of Fig.~\ref{fig:M1_T} (with $\alpha_i \sim 10^{-2}$). In spite of such low $M_i^0$, the $M_i(T_*)$ can easily satisfy $M_i(T_*)> T_*> T_{\rm sp}$ so that the decay of both the RHNs may take place at $T_*$. The mass-splitting between the two almost degenerate RHNs plays an important role in obtaining the large CP asymmetry, 
and is represented by $\Delta M(T)$ at $T > T_{\rm sp}$ as
\begin{align}
\Delta M (T) = \frac{1}{16}\text{Re}\left[(Y_\nu^\dagger Y_\nu)_{21}\right] \frac{T^2}{M(T)} + \alpha_{21}
A T^2,
\end{align}
where $\alpha_{21} = \alpha_2 - \alpha_1$. The first term corresponds to the thermally induced mass splitting ~\cite{Pilaftsis:2003gt} that being effectively $T$ independent, remains subdominant compared to the second one for sufficiently high $T_*$. The required mass splitting at $T_*$ can therefore be obtained by appropriately choosing $\alpha_{21}$ while parameter $A$ is  so chosen not to have $M_i(T)<T$ above $T_{\rm{sp}}$. 

On the other hand, due to the instant decay of the RHNs at $T_*$, radiation density may increase which can also dilute the produced lepton asymmetry to some extent. To incorporate all such effects during the production and subsequent evolution of the $B-L$ asymmetry, we solve the coupled Boltzmann equations (BE) for the number density of RHNs 
$n_{N_i}$, and $B-L$ asymmetry $n_{B-L}$ along with radiation energy density $\rho_R$ from $T=T_*$, which read as\footnote{Note that $\phi$ being in thermal equilibrium till the EW scale, will not contribute to the set of BE.} 
\begin{align}
&	\frac{d N_{i}}{d\mathcal{A}} = -(N_{i} - N_{i}^{eq})\frac{ \mathcal{K}_i}{\mathcal{A}}, ~~{\rm{with}}~~\mathcal{K}_i = \frac{\langle \Gamma_{N_{i}}\rangle}{H}, \label{eq:N}\\
&	\frac{dR}{d\mathcal{A}} =  \sum_{i=1}^{2}\mathcal{K}_i \left[M_{i}^{\rm{T}}(N_{i}-N_{i}^{eq})\right], \label{eq:r}\\
&	\frac{dN_{B-L}}{d\mathcal{A}} =  -\sum_{i=1}^{2}\frac{\mathcal{K}_i}{\mathcal{A}}\left[\varepsilon^{i}_\ell(N_{i}-N_{i}^{eq}) + \frac{n_{N_{i}}^{eq}}{2 n_{l}^{eq}}N_{B-L}\right]\label{eq:b-l}
\end{align}
where $\mathcal{A}=a/a_*$ is the normalized scale factor of the Universe with $a(T_*) = a_*$ satisfying $\mathcal{A}(T_*)=1$; $N_{i(B-L)}=n_{{N_{i}}(B-L)}\hspace{0.1cm}\mathcal{A}^3$; 
$R=\rho_{R}\hspace{0.1cm}\mathcal{A}^4$, $\langle\Gamma_{N_{i}}\rangle=\frac{K_1(\mathcal{A})}{K_2(\mathcal{A})}\Gamma_{N_i}$, Hubble parameter $\mathcal{H}=\sqrt{\frac{\rho_{R}\hspace{0.1cm}+\hspace{0.1cm}\sum_{i}M_{i}(T) n_{N_{i}}}{3 M_{P}^2}}$ and $M_i^{\rm{T}}$ represents the RHN mass inclusive of thermal correction at $T$ with $\rho_R (\mathcal{A}=1)=\frac{\pi^2}{30} g_* T_*^4$. Here, $g_*$ represents the relativistic degrees of freedom present in the thermal bath.

\begin{figure}[!htb]
\includegraphics[width=1\linewidth]{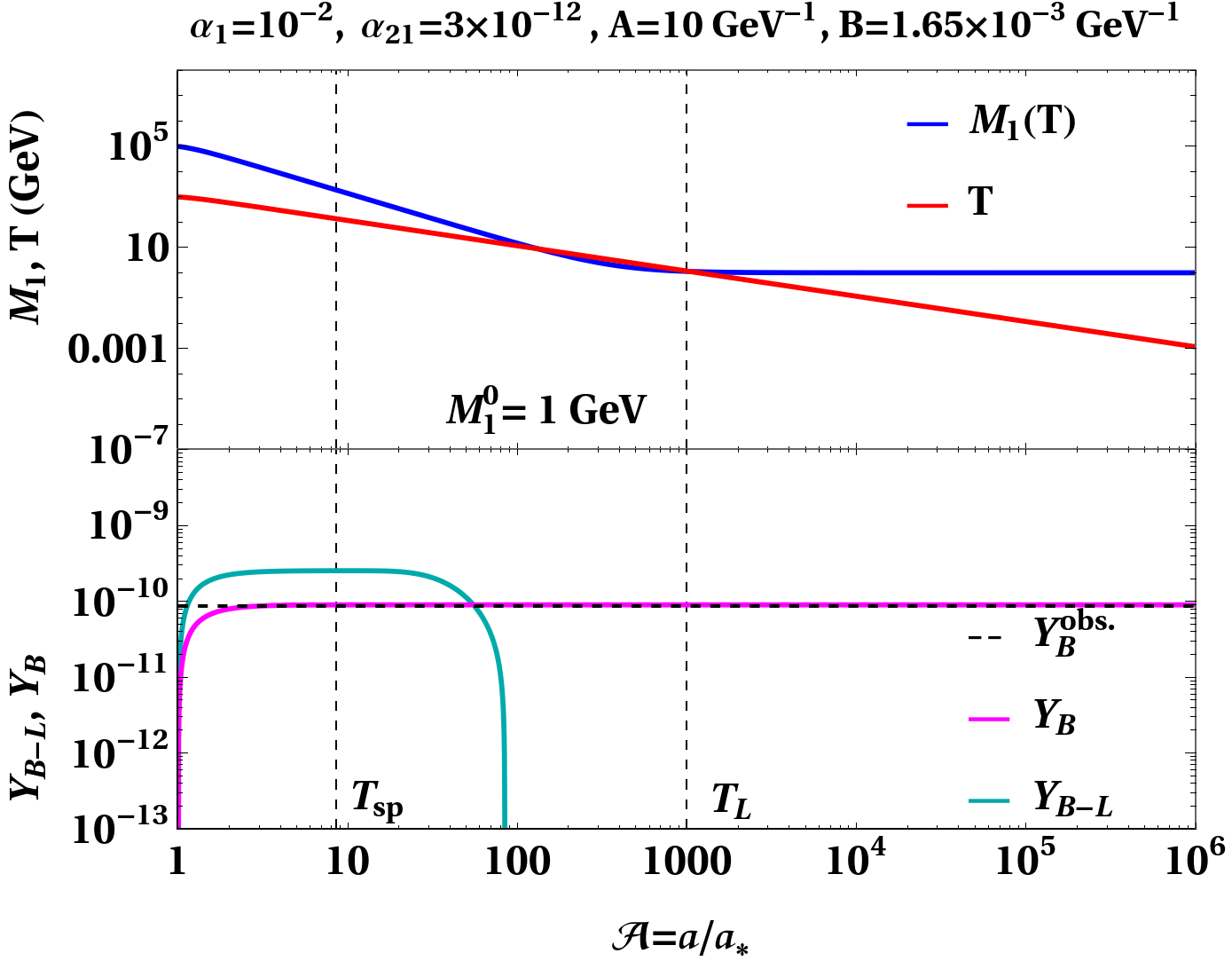}
\caption{ $M_1,~T$ (top panel) and $Y_{B-L},~Y_{B}$ (bottom panel)  variation against normalized scale factor $\mathcal{A}=a/a_*$ with $M_{1}^0=1$ GeV and ${\rm Re}(\theta)=0.78,~{\rm Im}(\theta)=-0.3$.}
\label{fig:yb-l}
\end{figure}

As stated earlier, prior to becoming massive at $T_*$, the RHNs are part of the thermal bath and hence  the moment it suddenly became massive, its number density can approximately be that of relativistic one at $\mathcal{A} = 1$. With such initial condition in mind, we solve Eqs.~\eqref{eq:N}-\eqref{eq:b-l} numerically. Fig.~\ref{fig:yb-l} shows the evolutions of $B-L$ asymmetry $Y_{B-L}$ (= $n_{B-L}/s$), baryon asymmetry $Y_B$  in the lower panel against $\mathcal{A}$ while $T$ and $M_i$ variations are depicted in upper panel. As seen, with  $T_*=1$ TeV, $\alpha_{21} = 3 \times 10^{-12}$ (corresponding to $\Delta M/M \sim 10^{-10}$), $A = 10~\text{GeV}^{-1} ,B=1.65 \times 10^{-3}~\text{ GeV}^{-1}$, a lepton asymmetry (cyan solid line) converted to requisite amount of baryon asymmetry (pink solid line) results with ZT RHN mass as low as 1 GeV.

As seen from Fig.~\ref{fig:yb-l}, after the sphaleron decoupling and baryon asymmetry freezes out, the temperature dependent RHN mass steadily decreases (as $\alpha_iBv^2(T)$ remains subdominant near $T_{\rm{c}}$) till a point beyond which it quickly saturates to its ZT constant mass $M_i^0$. As a result, at some temperature below $T_{\rm{c}}$, it becomes smaller than the surrounding temperature causing the inverse decay processes efficient to bring RHNs in thermal equilibrium, hence a sharp fall of lepton asymmetry is observed. It is interesting to note that due to this constant nature of final ZT masses of RHNs, there exists another phase (below EWSB) for which $M_i^0 > T$ happens at a late temperature marked by $T_{\rm{L}}$ in Fig.~\ref{fig:yb-l}. This particular $M_i^0$ being lighter than Higgs mass $m_h \sim 125$ GeV in this EW broken phase, no further decay of RHN is possible\footnote{These RHNs would decay~\cite{Pal:1981rm, Barger:1995ty,Boyarsky:2009ix,Atre:2009rg,Domcke:2020ety} before BBN (via active-sterile neutrino mixing).}. However, in case $M_{i}^0$ remains above the EW scale, this observation opens up an intriguing possibility in our set-up to  produce a primordial lepton asymmetry\footnote{Recently an attempt is made~\cite{ChoeJo:2023cnx} explaining such late $Y_L$ albeit RHNs are lighter than the SM Higgs.} from the late (after EWSB) decay into SM Higgs and lepton producing a late lepton asymmetry $Y_L^{\rm{late}}$. Such a non-zero lepton asymmetry may alter the primordial helium abundance ($Y_P$) \cite{ParticleDataGroup:2022pth,Matsumoto:2022tlr} 
as compared to the prediction of standard BBN ($Y_P^{\rm{BBN}}$) by a relation~\cite{Kawasaki:2022hvx,Escudero:2022okz} $Y_P \simeq Y_P^{\rm{BBN}} e^{-0.96\xi_{\nu_e}}$ where $\xi_{\nu_e}$ is the electron neutrino chemical potential bounded by $\xi_{\nu_e} = 0.008 \pm 0.013$ as per the PDG-21 data \cite{ParticleDataGroup:2022pth} connected to $Y_L$ via $Y_L=3 Y_{\nu_e}\simeq 0.105 \xi_{\nu_e}$. In case $\xi_{\nu_e}$ happens to be negative, thanks to the effectively temperature independent nature of the CP asymmetry parameter $\varepsilon_{\ell}$, there can be a direct correlation between $Y_L^{\rm{late}}$ and the baryon asymmetry via early leptogenesis in our proposal, a unique feature that is not explored so far to the best of our knowledge. 

We further note that in making $M_i^0 > m_h +m_{\ell}$ by increasing the parameter $B$ in order to realize the late decay of the RHNs, the $M_i(T)$ line (blue in Fig.~\ref{fig:yb-l}) shifts upward thereby increasing the tension 
with $M_i(T) < T$ realization after $T_{\rm{sp}}$, which is crucial for keeping the RHNs in equilibrium and
to be taken as the initial condition for late leptogenesis. As a heuristic approach, we find that with $[A, B] =[0.724, 1.78]~\text{GeV}^{-1}$ and $\alpha_1\sim 10^{-2},~\alpha_{21}=5.01 \times 10^{-15}$, it may lead to $Y_{L}^{\rm{late}} \sim 10^{-4}$ (corresponding to $|\xi_{\nu_e}| \sim 0.001$) where the CP asymmetry parameter $\varepsilon_{\ell}$, entering in both late and early resonant leptogenesis, turns out to be $\sim 0.06$. The early leptogenesis although consists of same  $\varepsilon_{\ell}$, the involvement of $T_*/M_i(T_*)$ in Eq.~\eqref{y_b-l}, needed to be $\mathcal{O}(10^{-9})$, provides a freedom to have the required $Y_{B-L}$.
This however poses a concern as such heavy RHNs while decaying at $T_*$ may increase the temperature of the Universe and demands a rigorous study, which is beyond the scope of the present analysis.
It can also be noticed that the recent EMPRESS result~\cite{Matsumoto:2022tlr} prefers a value of $\xi_{\nu_e}$ which is in the positive side and larger in magnitude and therefore difficult to reproduce in the present scenario. However there is a substantial uncertainty involved in determining $\xi_{\nu_e}$ precisely enough, making our scenario falsifiable in upcoming precision experiments like Simons Observatory \cite{SimonsObservatory:2018koc}, CMB-S4 \cite{Abazajian:2019eic}.

In summary, we have shown that a gradual settlement of $vev$ of a complex scalar field, coupled with RHNs, in the early Universe following a phase transition can give rise to a temperature dependent mass of RHNs which eventually affects 
the leptogenesis in a non-trivial way. The uniqueness of this scenario lies in the fact that not only did we obtain a viable leptogenesis scenario where instantaneous decay of $both$ the RHNs contributes to the lepton asymmetry resulting in a much lower bound on RHN mass compared to DI bound in the case with hierarchical RHNs, but also in case of quasi degenerate RHNs, leptogenesis remains viable for zero temperature RHN mass below the electroweak scale. This finding carries profound importance from the perspective of finding them in current and future facilities like in SHiP~\cite{Alekhin:2015byh}, FCC-ee \cite{Blondel:2014bra,Blondel:2022qqo} including their collider searches \cite{Antusch:2016vyf,Li:2023tbx}. 
Finally, we observe that the construction is capable of predicting a primordial lepton asymmetry, connected to early leptogenesis, in a somewhat tuned way which makes the framework testable in future experiments of precise determination of primordial helium abundance. 

\begin{acknowledgements}
The work of DB is supported by Council of Scientific \& Industrial Research (CSIR), Govt. of India, under the senior research fellowship scheme. AD acknowledges the the National Research Foundation of Korea (NRF) grant funded by the Korean government (2022R1A5A1030700) and the support provided by the Department of Physics, Kyungpook National University. The work of AS is supported by the grants CRG/2021/005080 and MTR/2021/000774 from SERB, Govt. of India. \end{acknowledgements}

\appendix

\onecolumngrid

\section{Vacuum structure of the relevant scalar fields and its Implications}\label{ap:1}
In this section, we provide a toy construction that effectively contributes to a temperature dependent (TD) dynamical vacuum expectation value ($vev$) for the $\phi$ field $v_{\phi}(T)$, hence providing a TD mass to right handed neutrinos (RHN) as in Eq. (3) of the main manuscript. For this purpose, we consider two additional Standard Model (SM) singlet scalar fields $S$ (carrying $B-L$ charge opposite to $\phi$ so as to form a neutral $\phi S$ combination) and $\eta$  which is neutral under $B-L$ symmetry. The relevant tree level zero temperature potential can be written as 
\begin{align}
	V_0(H,\phi,S,\eta)\supset   -\mu^2 H^\dagger H +\lambda(H^\dagger H)^2 +m_\phi^2 |\phi|^2
	+( A_1 \phi S  H^\dagger H +A_2 \phi S \eta^2 +~\text{h.c.}),
	\label{apeq:potential}
\end{align} 
where, $A_{1,2} <0$ are constant (real) couplings and $\mu^2>0$. Here we presume all other terms allowed by the symmetry are of negligible impact, and not included (in the limit of associated couplings to be zero) for simplicity. 
In the early Universe, with the high-temperature thermal bath comprising of all the SM particles along with the additional scalar fields $\phi$, $S$, and $\eta$, it is necessary to include thermal corrections to the above zero-temperature potential which, at the one-loop level, takes the form~~\cite{Dolan:1973qd, Quiros:1999jp,Batell:2022qvr} 
\begin{align}
	V_{T}(h,\phi,S,\eta ;T)  \supset & \frac{T^4}{2\pi^2}J_{B}\left(\frac{m_{h}^2(h,\phi,S,\eta,T)}{T^2}\right)
	+ \frac{3 T^4}{2\pi^2} J_{B}\left(\frac{m_{\chi}^2(h,\phi,S,\eta,T)}{T^2}\right)
	+ \sum_{i} g_{i}\frac{T^4}{2\pi^2}J_{B,F}\left(\frac{m_{i}^2(h,T)}{T^2}\right) \notag\\
	& + \frac{T^4}{2\pi^2}J_{B}\left(\frac{m_{\eta}^2(h,\phi,S,\eta,T)}{T^2}\right) +\sum_{j} g_{j}\frac{T^4}{2\pi^2}J_{F}\left(\frac{m_{j}^2(\phi,S)}{T^2}\right),	
	\label{eq4}
\end{align} 
where $J_{B,F}(z^2)=\int_{0}^{\infty} dx x^2 {\rm ln}[1\mp exp(-\sqrt{x^2+z^2})$ is the thermal function for bosons ($B$), fermions ($F$) and $m_{i,j}$ ($g_{i,j}$) denotes the field-dependent masses (degrees of freedom) for particles $i = [W_{T,L},Z_{T,L},A_{L};~t]$ and $j=[N_1,N_2]$ with $[W_{T,L},Z_{T,L},A_{L};t]$ representing transverse and longitudinal modes of three SM gauge bosons and top quark~\cite{Kang:2017mkl} and $[N_1,N_2]$ denote the RHNs respectively. Here, $h$ is the classical Higgs background and $\chi$ denotes the Nambu-Goldstone bosons of the SM Higgs. 

Before estimating the vacuum structure of the $\phi$ field, we comment on the $vev$s of the two additional fields, $S$ and $\eta$. While $\eta$ does not get any $vev$, we assume $S$ field (real part of it) obtains a $vev$ at a temperature $T_*$ so as to express it as
\begin{equation}
	\langle S \rangle=v_s =
	\begin{cases}
		0& \text{for} \quad T>T_*,\\
		v_1& \text{for}  \quad  T\leq T_*.
	\end{cases}
	\label{eq:vs}
\end{equation} 
This further enables us to determine the $vev$ for the $\phi$ field at $T_*$ and thereafter, via minimization of the potential (neglecting the zero temperature one-loop Coleman-Weinberg correction~\cite{Coleman:1973jx} for simplicity) $V = V_0 +V_T$ as we discuss now. Firstly, we note that the $vev$ of the SM Higgs is obtainable from the minimization condition $\frac{\partial V}{\partial h}=0$, which leads to electroweak phase transition (EWPT)
as a smooth standard crossover~\cite{Kajantie:1996mn} at $T = T_c ~(< T_*)$, given by
\begin{equation}
	\langle h \rangle\equiv	v(T) =
	\begin{cases}
		0& \text{for} \quad T>T_{c},\\
		\sqrt{\frac{\mu^2 - c_h T^2}{\lambda}} & \text{for}  \quad  T\leq T_{c},
	\end{cases}
	\label{eq:vh}
\end{equation}
where, $c_h=\left(\frac{\lambda}{2}+\frac{3}{16}g_{\rm{w}}^2+\frac{1}{16}g_{\rm{y}}^2+\frac{y_t^2}{4}\right)$; $g_{\rm{w}}, g_{\rm{y}}$ and $y_t$ are the $SU(2)_L, U(1)_Y$ gauge couplings and top quark Yukawa coupling respectively. It can be noted that the evolution of the SM Higgs vacuum remains unaffected by the presence of $\phi$ and $S$ provided $|A_1| v_{\phi} (T) v_s \ll \mu^2$ is assumed around $T_c$. 
Consequently, the $vev$ structure of $\phi$ field (only the real part) follows from $\frac{\partial V}{\partial \phi}=0$,
which leads to 
\begin{align}
	& \langle \phi\rangle \equiv v_\phi(T) \nonumber \\
	& = - \frac{v_s}{m_\phi^2} \left[
	\frac{A_1 T^2}{2\pi^2}\left\{J_{B}^\prime\left(\frac{m_{h}^2(h,\phi,S,\eta,T)}{T^2}\right)+3J_{B}^\prime\left(\frac{m_{\chi}^2(h,\phi,S,\eta,T)}{T^2}\right)\right\} 
	+ \frac{A_2 T^2}{\pi^2} J_{B}^\prime\left(\frac{m_{\eta}^2(h,\phi,S,\eta,T)}{T^2}\right) + \frac{A_1}{2} v^2(T) \right], \label{apeq:phi} \\
	& = A T^2 + B v^2(T),	\label{eq3}		 
\end{align}
where 
\begin{equation}
	A = - \frac{v_s}{m_\phi^2} \left[
	\frac{A_1}{2\pi^2}\left\{J_{B}^\prime\left(\frac{m_{h}^2(h,\phi,S,\eta,T)}{T^2}\right)+3J_{B}^\prime\left(\frac{m_{\chi}^2(h,\phi,S,\eta,T)}{T^2}\right)\right\} 
	+ \frac{A_2}{\pi^2} J_{B}^\prime\left(\frac{m_{\eta}^2(h,\phi,S,\eta,T)}{T^2}\right)\right]; ~B = - \frac{A_1 v_s}{2 m^2_{\phi}} ,
\end{equation} 
are constants (independent of temperature) as $J'_{B,F}(z^2)= \frac{\partial}{\partial z^2 } [J_{B,F}(z^2)]$ remains effectively temperature independent in leading order at high temperature expansion~\cite{Curtin:2016urg}. 
This exhibits the dependency of $v_{\phi}$ on temperature as we considered in the main manuscript. 

So what we could realize can be summed up in a nutshell here. In the early Universe with temperature above $T>T_*$, due to the vanishing $vev$ of the $S$ field, $v_\phi$ remains zero while for $ T_c < T \leq T_*$, the nonzero $v_s$ generates a 
temperature dependent $vev$ of $\phi$ as $v_{\phi}(T) = AT^2$ via Eq. \eqref{eq3} and thereafter once the EWPT initiates, $B$ term comes into the picture. This entire $T$ dependence of  $v_\phi$ is what stated in the manuscript. Note that due to the nonzero $v_s$ at $T<T_*$, RHN mass attains an additional contribution through the allowed interaction term $\alpha_{s_i} S^{\dagger}NN$. However with the condition $\alpha_{s_i} v_s<<\alpha_i B v^2$ our construction remains effectively  unaltered. Due to the involvement of free parameters such as $A_{1,2}, m_{\phi}, v_s$, a realization of values of $A$ and $B$ considered in the manuscript are achievable.
Though we have done the above analysis by retaining only the relevant terms (as in Eq.~\eqref{apeq:potential}) for simplicity, 
	a detailed analysis after including all the terms in the potential allowed by $U(1)_{B-L}$ symmetry would continue to provide a temperature dependent $vev$ profile for the $\phi$ field close to the Eq.~\eqref{eq3} (by adjusting the additional couplings), and thereby upholds the core assertion of our proposal.
	
	Finally, it is pertinent to discuss the fate of the additional scalar field in this setup.
	As discussed earlier, the scalar fields $\phi$, $S$, and $\eta$ should remain in thermal equilibrium till the EW scale so that the thermal corrections to the zero-temperature potential in Eq.~\eqref{eq4} can remain valid. Such consideration imposes constraints on their masses. Specifically, these scalar fields should preferably have masses below the EW scale ($\sim 160$ GeV) and must decay before nucleosynthesis in order not to disturb the big bang nucleosynthesis (BBN) predictions. The $\phi$ field can decay into $S$ and/or $\eta$ fields via its portal coupling provided $m_{\phi} > 2 m_{s,\eta}$. Given this mass constraint on $S$ and $\eta$, and to comply with the Higgs invisible decay limit, their masses need to lie within a range 62.5 GeV $< m_{s,\eta} < 80$ GeV. These scalars would eventually decay radiatively to SM leptons and quarks before BBN.

\twocolumngrid
\bibliography{ref.bib}

\end{document}